\documentclass[a4paper,10pt]{elsarticle}

\usepackage{graphicx}
\usepackage{subfigure}
\usepackage{amsmath, amsthm, amssymb}
\usepackage[utf8]{inputenc}
\usepackage{natbib}
\usepackage{hyperref}
\hypersetup{
    bookmarks=true,         
    pdffitwindow=false,     
    pdfstartview={FitH},    
    pdftitle={Growth Angle and Melt Meniscus of the RF-heated Floating Zone in Silicon Crystal Growth},    
    pdfauthor={Michael Wünscher, Anke Lüdge, Helge Riemann},     
    pdfcreator={Michael Wünscher},   
    pdfproducer={Michael Wünscher}, 
    pdfkeywords={Growth angle} {Free surface} {Laplace-Young equation} {Floating zone technique} {Growth from melt} {Semiconducting silicon}, 
    pdfnewwindow=true,      
    colorlinks=true,       
    linkcolor=red,          
    citecolor=green,        
    filecolor=magenta,      
    urlcolor=cyan           
}

\begin{document}
\begin{frontmatter}
\title{Growth Angle and Melt Meniscus of the RF-heated Floating Zone in Silicon Crystal Growth}

\author{Michael Wünscher}
\ead{wuenscher@ikz-berlin.de}
\author{Anke Lüdge}
\author{Helge Riemann}

\address{Leibniz Institute for Crystal Growth, Max-Born-Str. 2, 12489 Berlin, Germany}

\begin{abstract}
This article presents a direct measurement of the growth angle during the growth of a cylindrical 2`` silicon crystal using a radio-frequency heated floating zone process. From the high-resolution pictures taken during the process, this growth angle was evaluated to be $11^{\circ} \pm 2^{\circ}$. Furthermore,  the free surface of the melt was modeled using the Laplace-Young equation. This model has to include the electromagnetic pressure calculated by the surface ring currents approximation. The results were compared to the experimental free surface derived from video frames. It could be shown that the calculated free surface will only fit the experimentally determined one if the right growth angle is considered.
\end{abstract}


\end{frontmatter}

\section{Introduction}
\footnotetext{link to the original document: \href{http://dx.doi.org/10.1016/j.jcrysgro.2010.11.101}{10.1016/j.jcrysgro.2010.11.101}}
The floating zone (FZ) technique is used to grow silicon single crystals from polycrystalline feed rods (see ref.\cite{Coriell1977} for details). FZ silicon is mainly used for high-power electronics and is standing out against crucible-grown Cz silicon through the low oxygen content and high electron lifetime. A detailed description is found in ref.\cite{Amm2004}.

Investigating the free surface of the melt, one advantage of the FZ process is the good visibility. One can take photographs of nearly the full contour of the melt zone. For optically heated FZ furnaces, some investigations about the free surface can be found in ref.\cite{Tegetmeier1995} for silicon and GaAs or in ref.\cite{Shyy1995} for sapphire. In our case, videos are normally taken to control the growth process by measuring zone height and diameter of the crystal. In this paper, we used video frames to compare the numerical calculation of the free surface, based on the paper of Lie et al. \cite{Lie1990}, with the experiments.

The second task of the paper is the verification of the growth angle. There are some evaluations of the angle, e.g. in references \cite{Surek1975,Tegetmeier1995,Satunkin2003}. For the FZ process, this was described for processes in mirror furnaces. In our investigations we use radio-frequency (RF) heated FZ furnaces. Observing the FZ growth of silicon in such a furnace, the presence of a growth angle could be doubted. Differences to observations in mirror furnaces could have been explained by the electromagnetic pressure. However, careful analysis of the captured free surface as well as the high-resolution photographs, enabled the determination of the growth angle.

\section{Model for the free surface of the melt\label{par:freesurface}}
\begin{figure}[hbtp]
\centering{
\includegraphics[width=7cm]{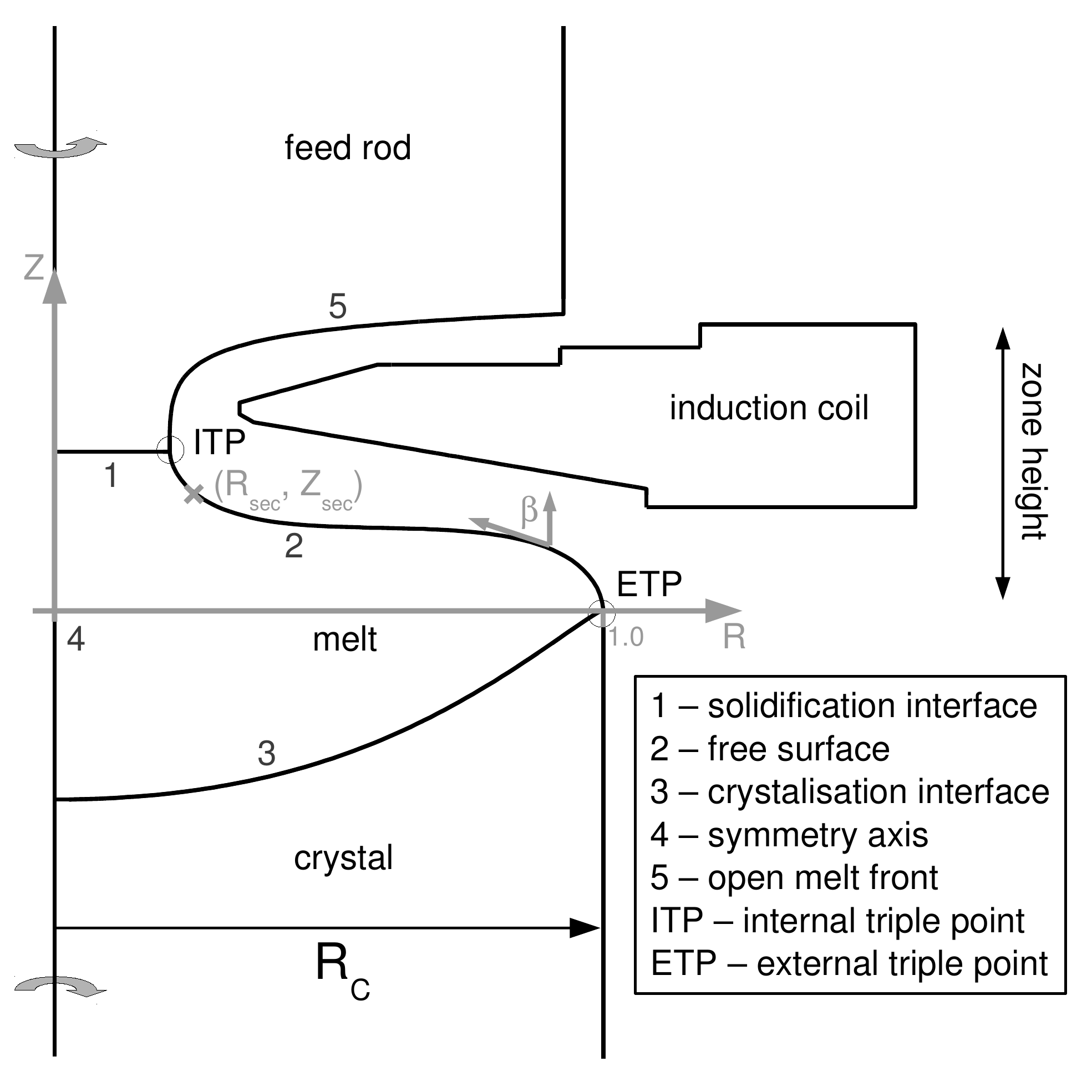}
\caption{Scheme of the FZ process, melt zone with the dimensionless coordinate system (grey) and the second point $(R_{sec},Z_{sec})$, if the solution fits.\label{fig:fzscematic}}
}
\end{figure}

The FZ process is nearly axisymmetric and, therefore, the pressure difference along the free surface of the melt was treated that way. The Laplace-Young equation \ref{eqn:lj} is solved on the geometry shown in fig. \ref{fig:fzscematic}.
\begin{equation}
    \Delta P-\rho gz + F_n = \gamma K\label{eqn:lj}\quad ,
 \end{equation}
with the pressure difference to surrounding gas $\Delta P$, the radial/vertical coordinates $r/z$, the electromagnetic pressure $F_n$, the mean curvature $K$, the surface tension $\gamma$, the density of the melt $\rho$ and the gravitational acceleration $g$. The equation is rewritten as a system of differential equations in order to be solved with numerical methods \cite{Lie1990}.

\begin{eqnarray}
	  \frac{dR}{ds} &=& sin \beta \\
	  \frac{dZ}{ds} &=& cos \beta \\
	  \frac{d\beta}{ds} &=& \frac{cos \beta}{R} + \underbrace{\frac{\rho g
R_C^2}{\gamma}}_{B_Z} Z +
\underbrace{\frac{V_0^2}{4\mu\gamma\omega^2R^3_C}}_{B_{El}} J^2_\theta(s) -
\underbrace{\Delta P \frac{R_C}{\gamma}}_{P_0}\label{eqn:lpydrei}
\end{eqnarray}
The dimensionless coordinates $R/Z$ are scaled with the crystal radius $R_C$ and the system has two material-dependent parameters, the Bond number $B_Z$ and the electrical Bond number $B_{El}$ depending on the electrical angular frequency $\omega$, the permeability $\mu$ and the applied voltage to the induction coil $V_0$.

This system of equations is solved by the Runge-Kutta method with a discretization step of 0.002 for the dimensionless arc length. We solve the system as an initial value problem with a start angle $\beta_0$ at position $(0.,1.)$ until curvature / angle at the end position are reached and the calculation is stopped. Typically the end angle is chosen to be zero, this means, that the curve ends vertically. 

The equation has one free parameter $P_0$, which is fixed by choosing a second point where the solution has to go through. This is accomplished by an optimization algorithm. From an initial solution, where $P_0$ is chosen to be not too far from the final solution, the distance to the present second point $(R_{sec},Z_{sec})$ is calculated. This point is determined by selecting the first point on the solution, which fits the radial coordinate $R_{sec}$. If the solution does not have such a point, the last point of the solution is selected. Therefore the last term of $P_0^{New}$ directs the solution towards the goal point. This term is zero for the last steps of convergence.
\begin{equation}
 P_0^{New} = P_0^{Old} - \frac{Z - Z_{sec}}{B_z}  - \frac{R - R_{sec}}{B_z}
\end{equation}
 
\begin{figure}[hbtp]
\centering{
	\includegraphics[width=7cm]{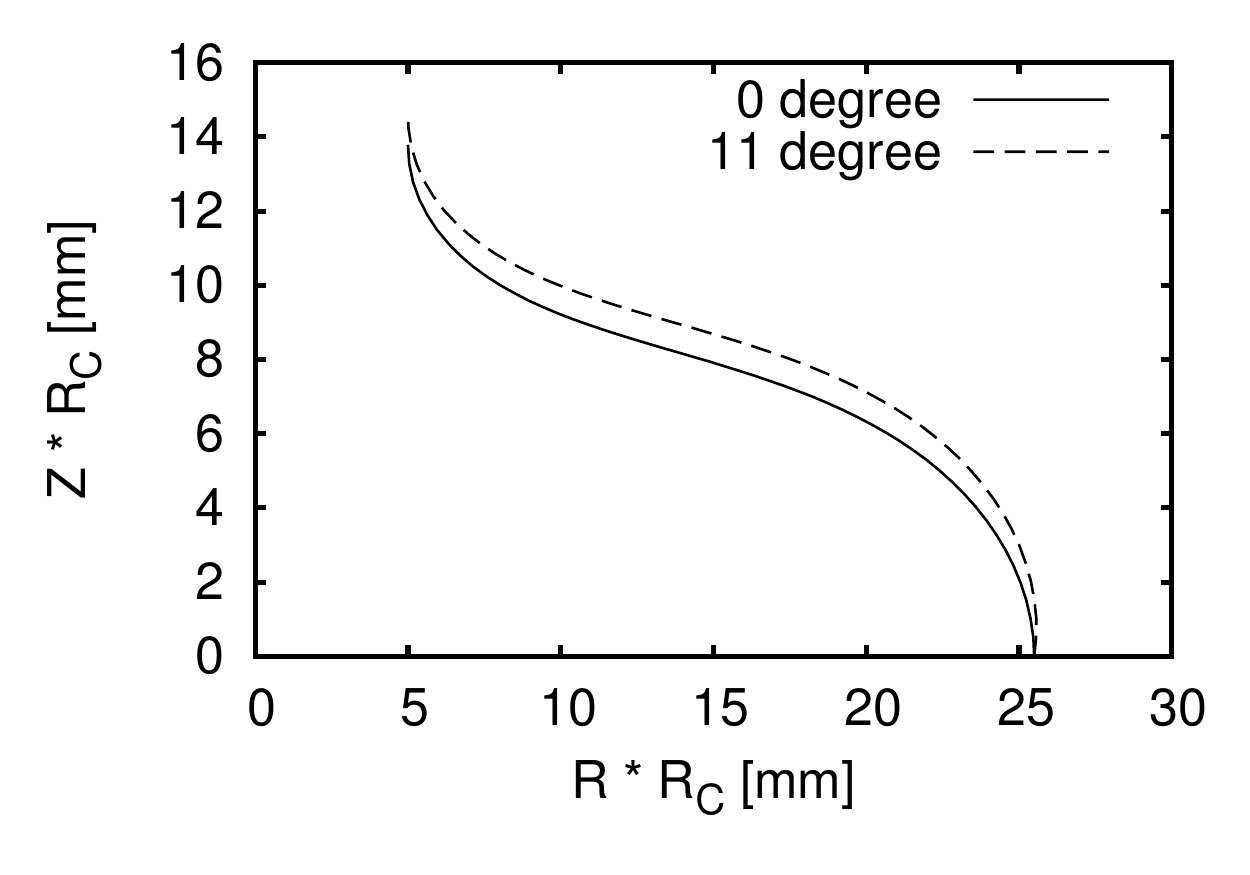}
}
\caption{Free surface calculation for different start angles ($0^\circ$ and $11^\circ$) without
electromagnetic pressure. Same neck diameter of 10mm with vertical ending for a 2'' FZ silicon process. \label{fig:diff_angle}}
\end{figure}

The growth angle is, as an initial condition for the solution, an important parameter for calculating the free surface. With a smaller start angle the solution has a lower melt height, which can be seen in figure \ref{fig:diff_angle}.

\section{Model for Electromagnetic Pressure Calculation}
The FZ process is heated by a RF induction coil at about 3MHz. From this coil a current is induced into the melt. As a side effect, there is the Lorentz force pressing against the free surface, which is taken into account with the third term of equation \ref{eqn:lpydrei} on the right-hand side. In this paragraph, we will show a way to calculate the current.

With respect to the skin depth $\delta = \sqrt{\frac{2}{\mu\omega\sigma}}\quad(\delta_{Cu,3MHz}\approx0.04mm)$ the current can be treated as surface current. For axisymmetry, it is further assumed that this surface current is modeled by ring currents \cite{Smythe1989} around the z-axis in cylindrical coordinates. With these assumptions, the dimensionless surface current $J_\theta$ is calculated as follows:
\begin{eqnarray}
 A_\theta &=& \frac{cos(\omega t)}{\pi} \int J_\theta(s) \sqrt{\frac{R}{mr}} [K(m)-E(m)]ds\label{eqn:integral}
\\
m&=&\frac{1-\sqrt{1-k^2}}{1+\sqrt{1-k^2}} \\
k^2&=&\frac{4rR}{(R+r)^2+(Z-z)^2}\quad ,
\end{eqnarray}
where R/Z are the coordinates of the ring current (source), which produces a magnetic field quantified by the vector potential at the coordinates r/z. All vector potentials are superimposed by integration along the surfaces of the conducting materials and give the equation for the system vector potential $A_\theta$ with the integration along the arc length $s$. In cylindrical coordinates this potential contains the functions $K(m)$ and $E(m)$, the complete elliptic integrals of first and second kind. Along the surfaces of crystal, melt and feed rod, where no voltage is applied, the vector potential $A_\theta$ is zero. Only along the induction coil, the high-frequency voltage gives a vector potential $A_\theta = \frac{cos(\omega t)}{2\pi r}$.

Together with the boundary conditions, we achieved an integral equation for the unknown surface current $J_\theta(s)$, which can be numerically solved by the boundary elements method. Therefore, the geometry is split into linear line segments numbered by $n$, where the dimensionless current $J_{\theta n}$ is assumed to be constant and equation \ref{eqn:integral}, only depending on the geometry, is solved with a 10-point Gaussian quadrature at the center of the line elements $(r_i,z_i)$, yielding the coefficients $A_{in}$.
\begin{equation}
 \sum^{\# segments}_{i=1}A_{in}J_{\theta n} =
\begin{cases}
(2\pi r_i)^{-1}&, \text{induction coil} \\
0&, \text{else}
\end{cases}
\end{equation}
The given system of linear equations, numbered by $i$, is solved with a standard direct numerical solver for the unknown currents $J_{\theta n}$. A detailed description can be found in the article of Lie et. al. \cite{Lie1990}. Our program is written in python using the scipy \cite{Jones2001} and numpy libraries as well as wxWidgets for the interface. A speed up of about 5-10 times is reached by using Fortran for calculating the integral.

By solving this linear system, we get the discrete form of the surface current $J_\theta(s)$. One numerical difficulty is to map this discrete solution to match the values required by the numerical solver of the Laplace-Young equation. Therefore we transform the numerical values from the cylindrical coordinate system into a parametrization along the arc length $s$. These values are then interpolated by a spline function of third order, which is used as input for the Runge-Kutta solver of the Laplace-Young equation in the \ref{par:freesurface}. paragraph. We toggle between both solvers until we do not have to change $P_0$ for a given error of $0.001$ in dimensionless units.  

\section{Comparison of numerical results with experiments}
To achieve pictures of the experiments, we used our camera system for measuring the diameter of the crystal during the growth of a 2`` crystal. These black and white picture have a resolution of $720 \times 576$ pixels what is $5.83$ pixel per mm. For a horizontally aligned camera, the inner end of the free surface is hidden by the induction coil. Therefore we cannot use the end of the surface as second point $(R_{sec},Z_{sec})$. Instead, we used an edge detection algorithm, which finds the point at 95\% of the visible height of the melt surface. It is not the last visible point because there are reflections of the melt at the induction coil, which make it impossible to find an accurate edge. The solution depends on the material parameter density $\varrho = 2530 \frac{kg}{m^3}$ and the surface tension $\gamma = 0.783 \frac{N}{m}$ \cite{Przyborowski1995}.
\begin{figure}[hbtp]
\centering{
	\subfigure[$\beta_0=0^{\circ}$ without EMP]{\includegraphics[width=3.9cm]{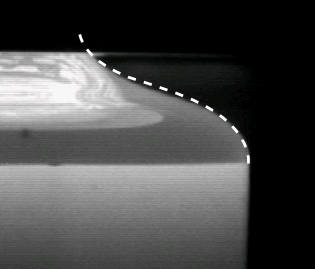}}
	\subfigure[$\beta_0=11^{\circ}$ without EMP]{\includegraphics[width=3.9cm]{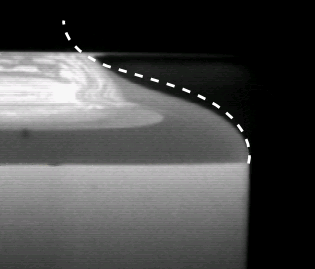}}
	\subfigure[$\beta_0=11^{\circ}$ with EMP]{\includegraphics[width=3.9cm]{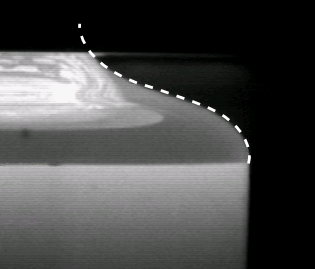}\label{fig:fitwithgood}}
}
\caption{Comparison of the numerical solution with the captured picture of the free surface. Electromagnetic pressure (EMP) $(V_0=500V)$ needs to be included to get a good fit with the experiment.\label{fig:fitwithout}}
\end{figure}

The first fits were made without including the electromagnetic pressure. Two solutions for initial angles of (a) $0^{\circ}$ and (b) $11^{\circ}$ are shown in figure \ref{fig:fitwithout}. Obviously, these solutions do not fit well. The deviation is especially large in the middle of the curve and even bigger for the solution with the right start angle. If the solution includes electromagnetic pressure of the right strength of 500V, the solution fits very well for the given resolution of the pictures (fig. \ref{fig:fitwithgood}).

As shown in the paper of Lie et al., the highest strength of the electromagnetic pressure is generated in the middle of the curve. This is also shown in our calculations. So the highest change of the free surface is achieved in this region. The electromagnetic pressure pushes the solution against the free surface. Of course, then one has to find a new $P_0$, else the second point would not lie on the curve anymore. 

\begin{figure}[hbtp]
\centering{
	\subfigure[$\beta_0=0^{\circ}$ ]{\includegraphics[width=3.9cm]{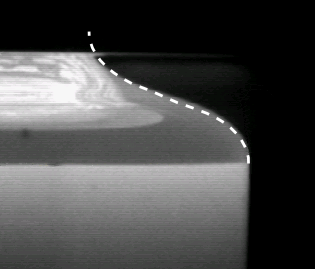}}
	\subfigure[$\beta_0=11^{\circ}$ ]{\includegraphics[width=3.9cm]{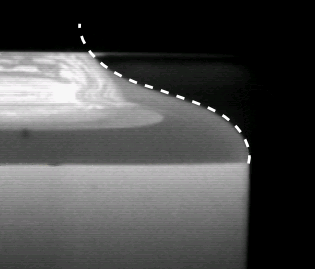}}
	\subfigure[$\beta_0=20^{\circ}$ ]{\includegraphics[width=3.9cm]{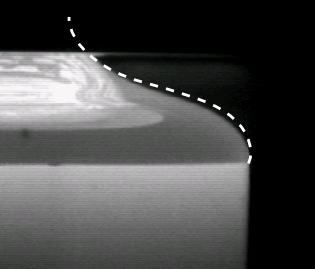}}
}
\caption{Different start angles with the same voltage $V_0=500V$ for the electromagnetic
pressure. Best equality with the experiment is given for a start angle of $11^\circ$, which is equal to the growth angle.\label{fig:diffgrowthangle}(published in \cite{Luedge2010})}
\end{figure}

With the same electromagnetic pressure, the start angle of $11^\circ$ shows the best fit to the experiment (see figure \ref{fig:diffgrowthangle}). So, the growth angle known from the literature could be verified. For the lowest angle (fig. \ref{fig:diffgrowthangle} a) the calculated solution is below the real melt surface and for the highest angle (fig. \ref{fig:diffgrowthangle} c) it is over it. It is very interesting to mention that the growth angle is not directly visible in the pictures because of the bad resolution, but this angle has a significant impact on the whole shape of the free surface. Therefore the growth angle is an important parameter to find the coincident free surface.

\begin{figure}[hbtp]
\centering{
	\subfigure[$\beta_0=9^{\circ}$ $V_0=400V$]{\includegraphics[width=3.9cm]{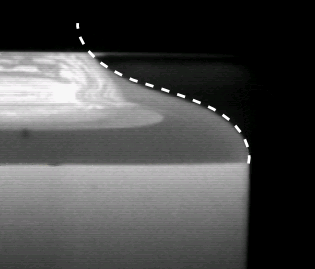}}
	\subfigure[$\beta_0=11^{\circ}$ $V_0=500V$]{\includegraphics[width=3.9cm]{w11_v500}}
	\subfigure[$\beta_0=13^{\circ}$ $V_0=600V$]{\includegraphics[width=3.9cm]{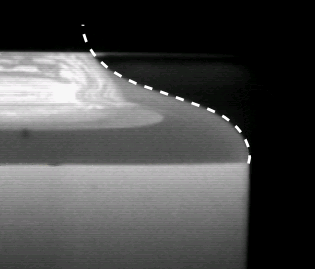}}
}
\caption{Best solution with different start angles $\beta_0$ and strengths $V_0$ of the electromagnetic pressure. Solution b) gives the best result. \label{fig:bestfit}}
\end{figure}

Nevertheless, in the applied way, this method has no better accuracy compared to a direct measurement of the growth angle as described later. In addition to $P_0$, the voltage for calculating the electromagnetic pressures is a second parameter, which has to be fitted. The dependency is shown in figure \ref{fig:bestfit}. Therefore one gets a parameter range for angle and voltage, where both parameters are not independent. Other uncertainties are geometry and position of the induction coil, which have to be modeled with a sufficient accuracy.

\section{Measurement of the growth angle from high-resolution photographs\label{sec:growthangle}}
As already mentioned, the growth angle is not directly visible in the cylindrical growth of the FZ crystals. Additionally, the small ripple of the crystal surface, which is caused by rotation, sometimes enhances this impression. Even looking at low-resolution video camera pictures does not show a visible angle, but evaluating high-resolution pictures from the experiment, this impression changes. 

We have used a Canon EOS 350 D with a Pentax 135mm lens mounted with three distance rings of together 50mm and an adapter Hama M42 to the camera. The equipment was put onto a tripod in front of the window in the growth chamber, which is around 350mm away from the growing crystal. It was separately tested that this equipment has a resolution of about 66 pixel per mm. Taking photos of the FZ process, it is not possible to use automatic focusing because there are many reflections. For taking good photographs we are focusing with a lower aperture, which has a lower depth of focus than a higher aperture. The maximum aperture was then used to take the photo at an exposure time of 1/160s. To reduce the impact of the lens distortion, we placed the area of interest in the middle of the photo.

\begin{figure}[hbtp]
\centering{
	\subfigure[Picture]{\includegraphics[width=5cm]{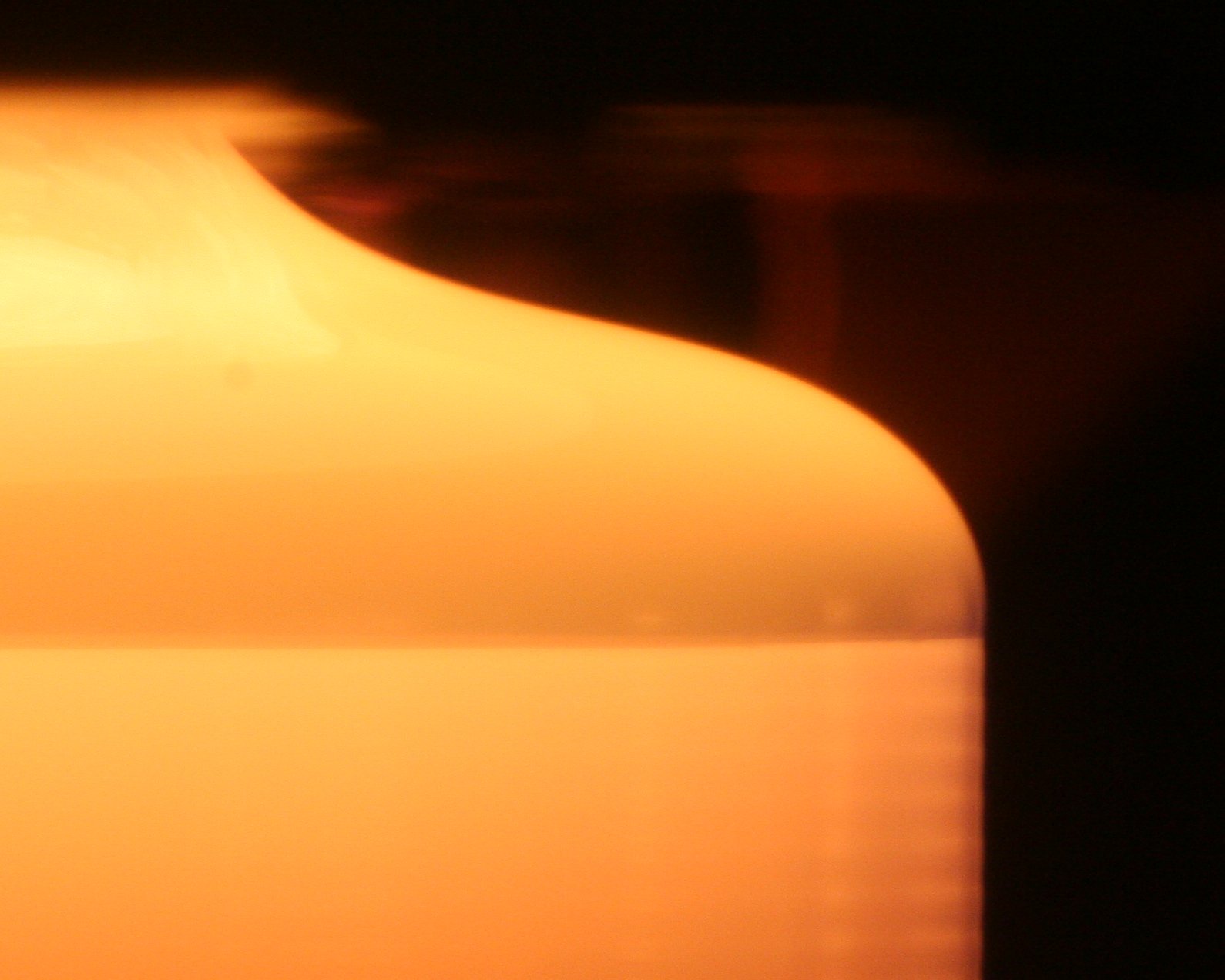}\label{fig:picangle}}
	\subfigure[Gradient]{\includegraphics[width=5cm]{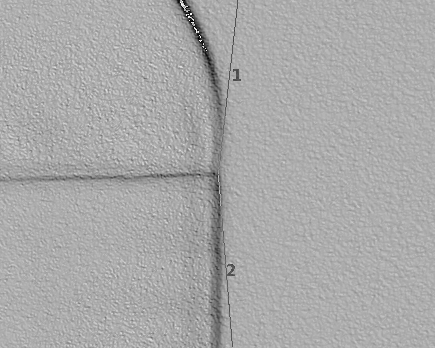}\label{fig:picanglegrad}}
}
\caption{(a) Picture of the growing 2'' silicon crystal (b) Gradient of the zoomed picture with
tangents to measure the growth angle $(11.1^{\circ})$}
\end{figure}

As in figure \ref{fig:picangle} the crystal radius is just reducing, so the outer shape is not vertical anymore. Therefore, the growth angle must not be measured against the vertical, but against the tangents of the crystal surface starting from the three-phase line. The edges get more visible from the Sobel gradient of the picture and two tangents are fitted (see \ref{fig:picanglegrad}). One tangent goes from the phase transition point along the free surface and the other one along the crystal. The angle between both tangents is the growth angle.

\begin{figure}[hbtp]
\centering{
	\subfigure[Begin of cylindrical growth]{\includegraphics[width=6cm]{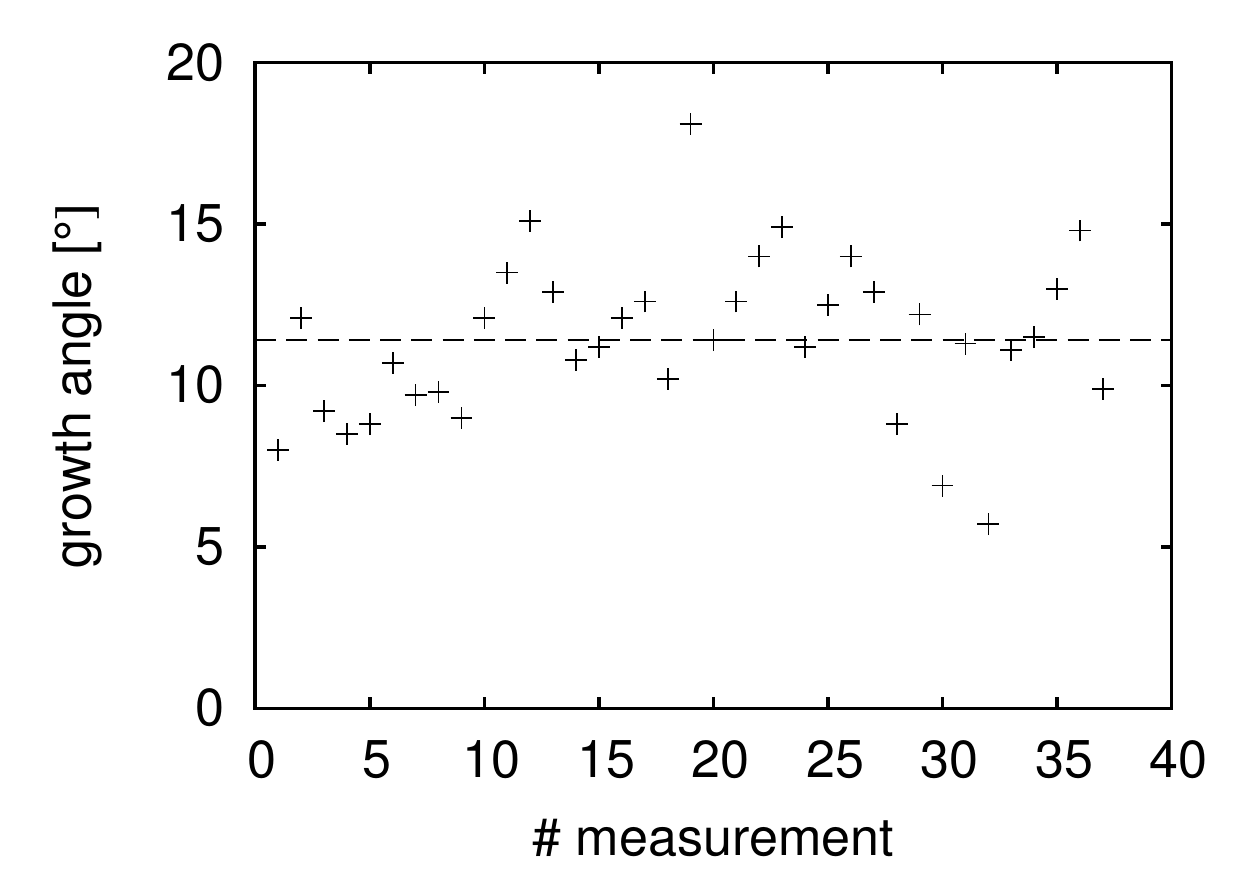}}
	\subfigure[End of cylindrical growth]{\includegraphics[width=6cm]{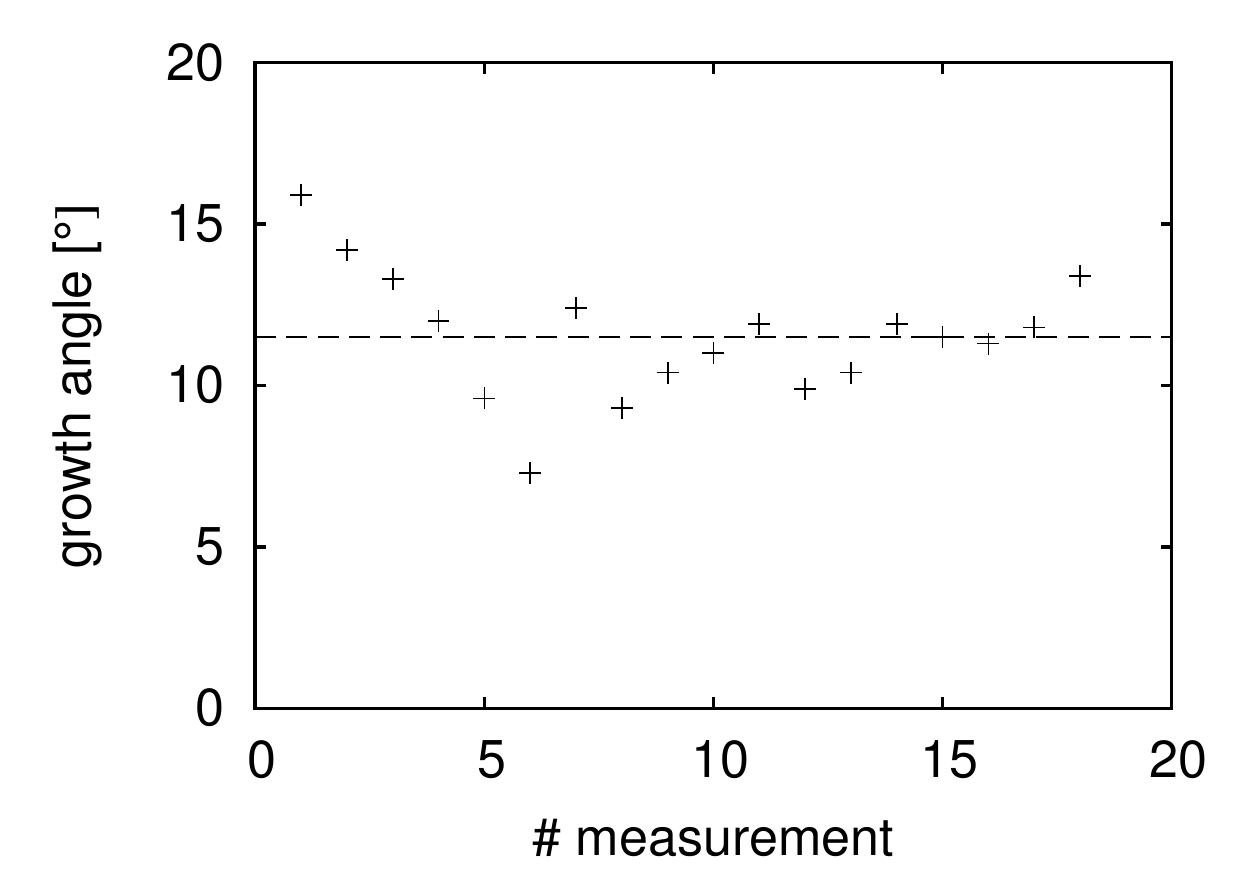}}
}
\caption{Measured growth angles with mean value (horizontal line) at two different stages of the growing crystal\label{fig:graphangle}}
\end{figure}

Because of the dynamics of the growth process, this angle is not constant. There are vibrations caused by rotation as well as jerks by the movement of crystal and feed rod. Also the three-phase line has some grooves at the facets, which break the rotational symmetry and will cause a small error in evaluation if the photo is taken nearby. During the experiment, we took several pictures similar to fig. \ref{fig:picangle} with a delay of some seconds, so we had 4-5 pictures per rotation. This was done at the beginning of the cylindrical growth and repeated after ten minutes. The growth angles measured from these photos are given in the graphs of fig. \ref{fig:graphangle}.
The mean growth angle is $11^{\circ} \pm 2^{\circ}$ for the first series and $11.5^{\circ} \pm 1.9^{\circ}$ for the last one. Theses values are in good agreement with the literature. In one work about the solidification of a wafer heated with an electron beam, the growth angle was found to be $11^{\circ}$ \cite{Surek1975}. A newer work \cite{Satunkin2003} of the solidification of a melt on top of a crystal found an equal result of $12^{\circ}$. For mirror furnaces, as mentioned before, Tegetmeier \cite{Tegetmeier1995} got a slightly different value of $8.6^{\circ}$ for silicon.

For a single evaluation of the growth angle, two angles need to be extracted from the gradient pictures. The reproducibility for that is better than $0.5^{\circ}$, therefore, the error for a single growth angle measurement is better than $1.0^{\circ}$. This contribution to the mean value is small and so we use the standard deviation as a good error estimate for our experiment. There are also some systematic errors which were tried to reduce. See experimental description above for details. 

\section{Conclusion}
We have achieved a good agreement of the simple model using the Laplace-Young equation together with the ring current approximation and the captured free surface of the experiment. For our fit we had to include the growth angle as initial condition as well as the effect of the electromagnetic pressure for the best results. Based on our good results we are looking forward to use our model to improve our control system for automatic crystal growth. It is also planned to increase the resolution of the camera system to get more measurement points along the free surface for a more precise comparison.

We have shown that the growth angle can be detected in RF-heated floating zone crystal growth and that it should be measured against the changing crystal surface.  Although this angle is hardly seen in our experiments, this angle was evaluated to be $11^{\circ} \pm 2^{\circ}$, which is in the range of the literature value of $11^{\circ}$. More accurate measurements could be achieved by a higher resolution of the camera, maybe with an automated system for the analyses, by reducing the movement of the melt by optimizing the growth parameters like pull speed, rotation rate or crystal radius, and by changing the lens to get a bigger depth of focus.

\bibliographystyle{elsarticle-num} 
\bibliography{article.bib}

\end{document}